\documentstyle[12pt,epsfig,a4,rotating]{article}
\newcommand{\charg}{{\chi}^{\pm}}

\def \epem {\mathrm{e^+ e^-}}

\def \mem  {\mathrm{MeV}/\it{c}^{\mathrm{2}}}
\def \gem  {\mathrm{GeV}/\it{c}^{\mathrm{2}}}

\def \gee  {\mathrm{GeV}}

\begin{document}
\date{}
\title{ \null\vspace{1cm}
Search for charginos nearly mass degenerate \\
  with the lightest neutralino in ${\rm e^+e^-}$ collisions \\
  at centre-of-mass energies up to 209\,GeV \\
\vspace{1cm}}
\author{The ALEPH Collaboration$^*)$}

\maketitle

\begin{picture}(160,1)
\put(0,115){\rm ORGANISATION EUROP\'EENNE POUR LA RECHERCHE
NUCL\'EAIRE (CERN)} \put(30,110){\rm Laboratoire Europ\'een pour
la Physique des Particules} \put(125,94){\parbox[t]{45mm}{\tt
CERN-EP/2002-020}} \put(125,88){\parbox[t]{45mm}{\tt 8 March 2002}}
\end{picture}

\vspace{.2cm}
\begin{abstract}
\vspace{.2cm} A search for charginos nearly mass degenerate with
the lightest neutralino is performed with the data collected by
the ALEPH detector at LEP, at centre-of-mass energies between 189
and 209\,GeV, corresponding to an integrated luminosity of
628\,pb$^{-1}$. The analysis is based on the detection of isolated
and energetic initial state radiation photons, produced in
association with chargino pairs whose decay products have little
visible energy. The number of candidate events observed is in
agreement with that expected from Standard Model background
sources. These results are combined with those of other
direct searches for charginos, and a lower limit of 88\,GeV/$c^2$
at 95\% confidence level is derived for the chargino mass in the
case of heavy sfermions, irrespective of the chargino-neutralino
mass difference.
\end{abstract}

\vfill
\centerline{\it Submitted to Physics Letters B}
\vskip .5cm
\noindent
--------------------------------------------\hfil\break
{\rm $^*)$ {\footnotesize{See next pages for the list of authors}} }

\eject


\pagestyle{empty}
\newpage
\small
%
%
\newlength{\saveparskip}
\newlength{\savetextheight}
\newlength{\savetopmargin}
\newlength{\savetextwidth}
\newlength{\saveoddsidemargin}
\newlength{\savetopsep}
\setlength{\saveparskip}{\parskip}
\setlength{\savetextheight}{\textheight}
\setlength{\savetopmargin}{\topmargin}
\setlength{\savetextwidth}{\textwidth}
\setlength{\saveoddsidemargin}{\oddsidemargin}
\setlength{\savetopsep}{\topsep}
%
%
\setlength{\parskip}{0.0cm}
\setlength{\textheight}{25.0cm}
\setlength{\topmargin}{-1.5cm}
\setlength{\textwidth}{16 cm}
\setlength{\oddsidemargin}{-0.0cm}
\setlength{\topsep}{1mm}
\pretolerance=10000
\centerline{\large\bf The ALEPH Collaboration}
\footnotesize
\vspace{0.5cm}
{\raggedbottom
\begin{sloppypar}
\samepage\noindent
A.~Heister,
S.~Schael
\nopagebreak
\begin{center}
\parbox{15.5cm}{\sl\samepage
Physikalisches Institut das RWTH-Aachen, D-52056 Aachen, Germany}
\end{center}\end{sloppypar}
\vspace{2mm}
\begin{sloppypar}
\noindent
R.~Barate,
R.~Bruneli\`ere,
I.~De~Bonis,
D.~Decamp,
C.~Goy,
S.~Jezequel,
J.-P.~Lees,
F.~Martin,
E.~Merle,
\mbox{M.-N.~Minard},
B.~Pietrzyk,
B.~Trocm\'e
\nopagebreak
\begin{center}
\parbox{15.5cm}{\sl\samepage
Laboratoire de Physique des Particules (LAPP), IN$^{2}$P$^{3}$-CNRS,
F-74019 Annecy-le-Vieux Cedex, France}
\end{center}\end{sloppypar}
\vspace{2mm}
\begin{sloppypar}
\noindent
G.~Boix,$^{25}$
S.~Bravo,
M.P.~Casado,
M.~Chmeissani,
J.M.~Crespo,
E.~Fernandez,
M.~Fernandez-Bosman,
Ll.~Garrido,$^{15}$
E.~Graug\'{e}s,
J.~Lopez,
M.~Martinez,
G.~Merino,
R.~Miquel,$^{4}$
Ll.M.~Mir,$^{4}$
A.~Pacheco,
D.~Paneque,
H.~Ruiz
\nopagebreak
\begin{center}
\parbox{15.5cm}{\sl\samepage
Institut de F\'{i}sica d'Altes Energies, Universitat Aut\`{o}noma
de Barcelona, E-08193 Bellaterra (Barcelona), Spain$^{7}$}
\end{center}\end{sloppypar}
\vspace{2mm}
\begin{sloppypar}
\noindent
A.~Colaleo,
D.~Creanza,
N.~De~Filippis,
M.~de~Palma,
G.~Iaselli,
G.~Maggi,
M.~Maggi,
S.~Nuzzo,
A.~Ranieri,
G.~Raso,$^{24}$
F.~Ruggieri,
G.~Selvaggi,
L.~Silvestris,
P.~Tempesta,
A.~Tricomi,$^{3}$
G.~Zito
\nopagebreak
\begin{center}
\parbox{15.5cm}{\sl\samepage
Dipartimento di Fisica, INFN Sezione di Bari, I-70126 Bari, Italy}
\end{center}\end{sloppypar}
\vspace{2mm}
\begin{sloppypar}
\noindent
X.~Huang,
J.~Lin,
Q. Ouyang,
T.~Wang,
Y.~Xie,
R.~Xu,
S.~Xue,
J.~Zhang,
L.~Zhang,
W.~Zhao
\nopagebreak
\begin{center}
\parbox{15.5cm}{\sl\samepage
Institute of High Energy Physics, Academia Sinica, Beijing, The People's
Republic of China$^{8}$}
\end{center}\end{sloppypar}
\vspace{2mm}
\begin{sloppypar}
\noindent
D.~Abbaneo,
P.~Azzurri,
T.~Barklow,$^{30}$
O.~Buchm\"uller,$^{30}$
M.~Cattaneo,
F.~Cerutti,
B.~Clerbaux,
H.~Drevermann,
R.W.~Forty,
M.~Frank,
F.~Gianotti,
T.C.~Greening,$^{26}$
J.B.~Hansen,
J.~Harvey,
D.E.~Hutchcroft,
P.~Janot,
B.~Jost,
M.~Kado,$^{2}$
P.~Mato,
A.~Moutoussi,
F.~Ranjard,
L.~Rolandi,
D.~Schlatter,
G.~Sguazzoni,
W.~Tejessy,
F.~Teubert,
A.~Valassi,
I.~Videau,
J.J.~Ward
\nopagebreak
\begin{center}
\parbox{15.5cm}{\sl\samepage
European Laboratory for Particle Physics (CERN), CH-1211 Geneva 23,
Switzerland}
\end{center}\end{sloppypar}
\vspace{2mm}
\begin{sloppypar}
\noindent
F.~Badaud,
S.~Dessagne,
A.~Falvard,$^{20}$
D.~Fayolle,
P.~Gay,
J.~Jousset,
B.~Michel,
S.~Monteil,
D.~Pallin,
J.M.~Pascolo,
P.~Perret
\nopagebreak
\begin{center}
\parbox{15.5cm}{\sl\samepage
Laboratoire de Physique Corpusculaire, Universit\'e Blaise Pascal,
IN$^{2}$P$^{3}$-CNRS, Clermont-Ferrand, F-63177 Aubi\`{e}re, France}
\end{center}\end{sloppypar}
\vspace{2mm}
\begin{sloppypar}
\noindent
J.D.~Hansen,
J.R.~Hansen,
P.H.~Hansen,
B.S.~Nilsson
\nopagebreak
\begin{center}
\parbox{15.5cm}{\sl\samepage
Niels Bohr Institute, 2100 Copenhagen, DK-Denmark$^{9}$}
\end{center}\end{sloppypar}
\vspace{2mm}
\begin{sloppypar}
\noindent
A.~Kyriakis,
C.~Markou,
E.~Simopoulou,
A.~Vayaki,
K.~Zachariadou
\nopagebreak
\begin{center}
\parbox{15.5cm}{\sl\samepage
Nuclear Research Center Demokritos (NRCD), GR-15310 Attiki, Greece}
\end{center}\end{sloppypar}
\vspace{2mm}
\begin{sloppypar}
\noindent
A.~Blondel,$^{12}$
\mbox{J.-C.~Brient},
F.~Machefert,
A.~Roug\'{e},
M.~Swynghedauw,
R.~Tanaka
\linebreak
H.~Videau
\nopagebreak
\begin{center}
\parbox{15.5cm}{\sl\samepage
Laboratoire de Physique Nucl\'eaire et des Hautes Energies, Ecole
Polytechnique, IN$^{2}$P$^{3}$-CNRS, \mbox{F-91128} Palaiseau Cedex, France}
\end{center}\end{sloppypar}
\vspace{2mm}
\begin{sloppypar}
\noindent
V.~Ciulli,
E.~Focardi,
G.~Parrini
\nopagebreak
\begin{center}
\parbox{15.5cm}{\sl\samepage
Dipartimento di Fisica, Universit\`a di Firenze, INFN Sezione di Firenze,
I-50125 Firenze, Italy}
\end{center}\end{sloppypar}
\vspace{2mm}
\begin{sloppypar}
\noindent
A.~Antonelli,
M.~Antonelli,
G.~Bencivenni,
F.~Bossi,
P.~Campana,
G.~Capon,
V.~Chiarella,
P.~Laurelli,
G.~Mannocchi,$^{5}$
F.~Murtas,
G.P.~Murtas,
L.~Passalacqua
\nopagebreak
\begin{center}
\parbox{15.5cm}{\sl\samepage
Laboratori Nazionali dell'INFN (LNF-INFN), I-00044 Frascati, Italy}
\end{center}\end{sloppypar}
\vspace{2mm}
\begin{sloppypar}
\noindent
A.~Halley,
J.~Kennedy,
J.G.~Lynch,
P.~Negus,
V.~O'Shea,
A.S.~Thompson
\nopagebreak
\begin{center}
\parbox{15.5cm}{\sl\samepage
Department of Physics and Astronomy, University of Glasgow, Glasgow G12
8QQ,United Kingdom$^{10}$}
\end{center}\end{sloppypar}
\vspace{2mm}
\pagebreak
\begin{sloppypar}
\noindent
S.~Wasserbaech
\nopagebreak
\begin{center}
\parbox{15.5cm}{\sl\samepage
Department of Physics, Haverford College, Haverford, PA 19041-1392, U.S.A.}
\end{center}\end{sloppypar}
\vspace{2mm}
\begin{sloppypar}
\noindent
R.~Cavanaugh,$^{33}$
S.~Dhamotharan,$^{34}$
C.~Geweniger,
P.~Hanke,
V.~Hepp,
E.E.~Kluge,
G.~Leibenguth,
A.~Putzer,
H.~Stenzel,
K.~Tittel,
M.~Wunsch$^{19}$
\nopagebreak
\begin{center}
\parbox{15.5cm}{\sl\samepage
Kirchhoff-Institut f\"ur Physik, Universit\"at Heidelberg, D-69120
Heidelberg, Germany$^{16}$}
\end{center}\end{sloppypar}
\vspace{2mm}
\begin{sloppypar}
\noindent
R.~Beuselinck,
W.~Cameron,
G.~Davies,
P.J.~Dornan,
M.~Girone,$^{1}$
R.D.~Hill,
N.~Marinelli,
J.~Nowell,
S.A.~Rutherford,
J.K.~Sedgbeer,
J.C.~Thompson,$^{14}$
R.~White
\nopagebreak
\begin{center}
\parbox{15.5cm}{\sl\samepage
Department of Physics, Imperial College, London SW7 2BZ,
United Kingdom$^{10}$}
\end{center}\end{sloppypar}
\vspace{2mm}
\begin{sloppypar}
\noindent
V.M.~Ghete,
P.~Girtler,
E.~Kneringer,
D.~Kuhn,
G.~Rudolph
\nopagebreak
\begin{center}
\parbox{15.5cm}{\sl\samepage
Institut f\"ur Experimentalphysik, Universit\"at Innsbruck, A-6020
Innsbruck, Austria$^{18}$}
\end{center}\end{sloppypar}
\vspace{2mm}
\begin{sloppypar}
\noindent
E.~Bouhova-Thacker,
C.K.~Bowdery,
D.P.~Clarke,
G.~Ellis,
A.J.~Finch,
F.~Foster,
G.~Hughes,
R.W.L.~Jones,
M.R.~Pearson,
N.A.~Robertson,
M.~Smizanska
\nopagebreak
\begin{center}
\parbox{15.5cm}{\sl\samepage
Department of Physics, University of Lancaster, Lancaster LA1 4YB,
United Kingdom$^{10}$}
\end{center}\end{sloppypar}
\vspace{2mm}
\begin{sloppypar}
\noindent
O.~van~der~Aa,
C.~Delaere,
V.~Lemaitre
\nopagebreak
\begin{center}
\parbox{15.5cm}{\sl\samepage
Institut de Physique Nucl\'eaire, D\'epartement de Physique, Universit\'e Catholique de Louvain, 1348 Louvain-la-Neuve, Belgium}
\end{center}\end{sloppypar}
\vspace{2mm}
\begin{sloppypar}
\noindent
U.~Blumenschein,
F.~H\"olldorfer,
K.~Jakobs,
F.~Kayser,
K.~Kleinknecht,
A.-S.~M\"uller,
G.~Quast,$^{6}$
B.~Renk,
H.-G.~Sander,
S.~Schmeling,
H.~Wachsmuth,
C.~Zeitnitz,
T.~Ziegler
\nopagebreak
\begin{center}
\parbox{15.5cm}{\sl\samepage
Institut f\"ur Physik, Universit\"at Mainz, D-55099 Mainz, Germany$^{16}$}
\end{center}\end{sloppypar}
\vspace{2mm}
\begin{sloppypar}
\noindent
A.~Bonissent,
P.~Coyle,
C.~Curtil,
A.~Ealet,
D.~Fouchez,
P.~Payre,
A.~Tilquin
\nopagebreak
\begin{center}
\parbox{15.5cm}{\sl\samepage
Centre de Physique des Particules de Marseille, Univ M\'editerran\'ee,
IN$^{2}$P$^{3}$-CNRS, F-13288 Marseille, France}
\end{center}\end{sloppypar}
\vspace{2mm}
\begin{sloppypar}
\noindent
F.~Ragusa
\nopagebreak
\begin{center}
\parbox{15.5cm}{\sl\samepage
Dipartimento di Fisica, Universit\`a di Milano e INFN Sezione di
Milano, I-20133 Milano, Italy.}
\end{center}\end{sloppypar}
\vspace{2mm}
\begin{sloppypar}
\noindent
A.~David,
H.~Dietl,
G.~Ganis,$^{27}$
K.~H\"uttmann,
G.~L\"utjens,
W.~M\"anner,
\mbox{H.-G.~Moser},
R.~Settles,
G.~Wolf
\nopagebreak
\begin{center}
\parbox{15.5cm}{\sl\samepage
Max-Planck-Institut f\"ur Physik, Werner-Heisenberg-Institut,
D-80805 M\"unchen, Germany\footnotemark[16]}
\end{center}\end{sloppypar}
\vspace{2mm}
\begin{sloppypar}
\noindent
J.~Boucrot,
O.~Callot,
M.~Davier,
L.~Duflot,
\mbox{J.-F.~Grivaz},
Ph.~Heusse,
A.~Jacholkowska,$^{32}$
C.~Loomis,
L.~Serin,
\mbox{J.-J.~Veillet},
J.-B.~de~Vivie~de~R\'egie,$^{28}$
C.~Yuan
\nopagebreak
\begin{center}
\parbox{15.5cm}{\sl\samepage
Laboratoire de l'Acc\'el\'erateur Lin\'eaire, Universit\'e de Paris-Sud,
IN$^{2}$P$^{3}$-CNRS, F-91898 Orsay Cedex, France}
\end{center}\end{sloppypar}
\vspace{2mm}
\begin{sloppypar}
\noindent
G.~Bagliesi,
T.~Boccali,
L.~Fo\`a,
A.~Giammanco,
A.~Giassi,
F.~Ligabue,
A.~Messineo,
F.~Palla,
G.~Sanguinetti,
A.~Sciab\`a,
R.~Tenchini,$^{1}$
A.~Venturi,$^{1}$
P.G.~Verdini
\samepage
\begin{center}
\parbox{15.5cm}{\sl\samepage
Dipartimento di Fisica dell'Universit\`a, INFN Sezione di Pisa,
e Scuola Normale Superiore, I-56010 Pisa, Italy}
\end{center}\end{sloppypar}
\vspace{2mm}
\begin{sloppypar}
\noindent
O.~Awunor,
G.A.~Blair,
G.~Cowan,
A.~Garcia-Bellido,
M.G.~Green,
L.T.~Jones,
T.~Medcalf,
A.~Misiejuk,
J.A.~Strong,
P.~Teixeira-Dias
\nopagebreak
\begin{center}
\parbox{15.5cm}{\sl\samepage
Department of Physics, Royal Holloway \& Bedford New College,
University of London, Egham, Surrey TW20 OEX, United Kingdom$^{10}$}
\end{center}\end{sloppypar}
\vspace{2mm}
\begin{sloppypar}
\noindent
R.W.~Clifft,
T.R.~Edgecock,
P.R.~Norton,
I.R.~Tomalin
\nopagebreak
\begin{center}
\parbox{15.5cm}{\sl\samepage
Particle Physics Dept., Rutherford Appleton Laboratory,
Chilton, Didcot, Oxon OX11 OQX, United Kingdom$^{10}$}
\end{center}\end{sloppypar}
\vspace{2mm}
\begin{sloppypar}
\noindent
\mbox{B.~Bloch-Devaux},
D.~Boumediene,
P.~Colas,
B.~Fabbro,
E.~Lan\c{c}on,
\mbox{M.-C.~Lemaire},
E.~Locci,
P.~Perez,
J.~Rander,
\mbox{J.-F.~Renardy},
A.~Rosowsky,
P.~Seager,$^{13}$
A.~Trabelsi,$^{21}$
B.~Tuchming,
B.~Vallage
\nopagebreak
\begin{center}
\parbox{15.5cm}{\sl\samepage
CEA, DAPNIA/Service de Physique des Particules,
CE-Saclay, F-91191 Gif-sur-Yvette Cedex, France$^{17}$}
\end{center}\end{sloppypar}
\vspace{2mm}
\begin{sloppypar}
\noindent
N.~Konstantinidis,
A.M.~Litke,
G.~Taylor
\nopagebreak
\begin{center}
\parbox{15.5cm}{\sl\samepage
Institute for Particle Physics, University of California at
Santa Cruz, Santa Cruz, CA 95064, USA$^{22}$}
\end{center}\end{sloppypar}
\vspace{2mm}
\begin{sloppypar}
\noindent
C.N.~Booth,
S.~Cartwright,
F.~Combley,$^{31}$
P.N.~Hodgson,
M.~Lehto,
L.F.~Thompson
\nopagebreak
\begin{center}
\parbox{15.5cm}{\sl\samepage
Department of Physics, University of Sheffield, Sheffield S3 7RH,
United Kingdom$^{10}$}
\end{center}\end{sloppypar}
\vspace{2mm}
\begin{sloppypar}
\noindent
K.~Affholderbach,$^{23}$
A.~B\"ohrer,
S.~Brandt,
C.~Grupen,
J.~Hess,
A.~Ngac,
G.~Prange,
U.~Sieler
\nopagebreak
\begin{center}
\parbox{15.5cm}{\sl\samepage
Fachbereich Physik, Universit\"at Siegen, D-57068 Siegen, Germany$^{16}$}
\end{center}\end{sloppypar}
\vspace{2mm}
\begin{sloppypar}
\noindent
C.~Borean,
G.~Giannini
\nopagebreak
\begin{center}
\parbox{15.5cm}{\sl\samepage
Dipartimento di Fisica, Universit\`a di Trieste e INFN Sezione di Trieste,
I-34127 Trieste, Italy}
\end{center}\end{sloppypar}
\vspace{2mm}
\begin{sloppypar}
\noindent
H.~He,
J.~Putz,
J.~Rothberg
\nopagebreak
\begin{center}
\parbox{15.5cm}{\sl\samepage
Experimental Elementary Particle Physics, University of Washington, Seattle,
WA 98195 U.S.A.}
\end{center}\end{sloppypar}
\vspace{2mm}
\begin{sloppypar}
\noindent
S.R.~Armstrong,
K.~Berkelman,
K.~Cranmer,
D.P.S.~Ferguson,
Y.~Gao,$^{29}$
S.~Gonz\'{a}lez,
O.J.~Hayes,
H.~Hu,
S.~Jin,
J.~Kile,
P.A.~McNamara III,
J.~Nielsen,
Y.B.~Pan,
\mbox{J.H.~von~Wimmersperg-Toeller}, 
W.~Wiedenmann,
J.~Wu,
Sau~Lan~Wu,
X.~Wu,
G.~Zobernig
\nopagebreak
\begin{center}
\parbox{15.5cm}{\sl\samepage
Department of Physics, University of Wisconsin, Madison, WI 53706,
USA$^{11}$}
\end{center}\end{sloppypar}
\vspace{2mm}
\begin{sloppypar}
\noindent
G.~Dissertori
\nopagebreak
\begin{center}
\parbox{15.5cm}{\sl\samepage
Institute for Particle Physics, ETH H\"onggerberg, 8093 Z\"urich,
Switzerland.}
\end{center}\end{sloppypar}
}
\footnotetext[1]{Also at CERN, 1211 Geneva 23, Switzerland.}
\footnotetext[2]{Now at Fermilab, PO Box 500, MS 352, Batavia, IL 60510, USA}
\footnotetext[3]{Also at Dipartimento di Fisica di Catania and INFN Sezione di
 Catania, 95129 Catania, Italy.}
\footnotetext[4]{Now at LBNL, Berkeley, CA 94720, U.S.A.}
\footnotetext[5]{Also Istituto di Cosmo-Geofisica del C.N.R., Torino,
Italy.}
\footnotetext[6]{Now at Institut f\"ur Experimentelle Kernphysik, Universit\"at Karlsruhe, 76128 Karlsruhe, Germany.}
\footnotetext[7]{Supported by CICYT, Spain.}
\footnotetext[8]{Supported by the National Science Foundation of China.}
\footnotetext[9]{Supported by the Danish Natural Science Research Council.}
\footnotetext[10]{Supported by the UK Particle Physics and Astronomy Research
Council.}
\footnotetext[11]{Supported by the US Department of Energy, grant
DE-FG0295-ER40896.}
\footnotetext[12]{Now at Departement de Physique Corpusculaire, Universit\'e de
Gen\`eve, 1211 Gen\`eve 4, Switzerland.}
\footnotetext[13]{Supported by the Commission of the European Communities,
contract ERBFMBICT982874.}
\footnotetext[14]{Supported by the Leverhulme Trust.}
\footnotetext[15]{Permanent address: Universitat de Barcelona, 08208 Barcelona,
Spain.}
\footnotetext[16]{Supported by Bundesministerium f\"ur Bildung
und Forschung, Germany.}
\footnotetext[17]{Supported by the Direction des Sciences de la
Mati\`ere, C.E.A.}
\footnotetext[18]{Supported by the Austrian Ministry for Science and Transport.}
\footnotetext[19]{Now at SAP AG, 69185 Walldorf, Germany}
\footnotetext[20]{Now at Groupe d' Astroparticules de Montpellier, Universit\'e de Montpellier II, 34095 Montpellier, France.}
\footnotetext[21]{Now at D\'epartement de Physique, Facult\'e des Sciences de Tunis, 1060 Le Belv\'ed\`ere, Tunisia.}
\footnotetext[22]{Supported by the US Department of Energy,
grant DE-FG03-92ER40689.}
\footnotetext[23]{Now at Skyguide, Swissair Navigation Services, Geneva, Switzerland.}
\footnotetext[24]{Also at Dipartimento di Fisica e Tecnologie Relative, Universit\`a di Palermo, Palermo, Italy.}
\footnotetext[25]{Now at McKinsey and Compagny, Avenue Louis Casal 18, 1203 Geneva, Switzerland.}
\footnotetext[26]{Now at Honeywell, Phoenix AZ, U.S.A.}
\footnotetext[27]{Now at INFN Sezione di Roma II, Dipartimento di Fisica, Universit\`a di Roma Tor Vergata, 00133 Roma, Italy.}
\footnotetext[28]{Now at Centre de Physique des Particules de Marseille, Univ M\'editerran\'ee, F-13288 Marseille, France.}
\footnotetext[29]{Also at Department of Physics, Tsinghua University, Beijing, The People's Republic of China.}
\footnotetext[30]{Now at SLAC, Stanford, CA 94309, U.S.A.}
\footnotetext[31]{Deceased.}
\footnotetext[32]{Also at Groupe d' Astroparticules de Montpellier, Universit\'e de Montpellier II, 34095 Montpellier, France.}  
\footnotetext[33]{Now at University of Florida, Department of Physics, Gainesville, Florida 32611-8440, USA}
\footnotetext[34]{Now at BNP Paribas, 60325 Frankfurt am Mainz, Germany}
\setlength{\parskip}{\saveparskip}
\setlength{\textheight}{\savetextheight}
\setlength{\topmargin}{\savetopmargin}
\setlength{\textwidth}{\savetextwidth}
\setlength{\oddsidemargin}{\saveoddsidemargin}
\setlength{\topsep}{\savetopsep}
\normalsize
\newpage
\pagestyle{plain}
\setcounter{page}{1}


\section{Introduction}
\label{sec:intro}
During its last three years of running (1998--2000), the ALEPH
detector at LEP collected data from ${\rm e}^+{\rm e}^-$
collisions at centre-of-mass energies between 189 and 209\,GeV,
corresponding to an integrated luminosity of 628\,${\rm pb}^{-1}$
(Table~1). In this letter, a search is performed on these data for
the pair production of charginos, ${\rm e}^+{\rm e}^- \to
\chi^+\chi^-$, in the framework of supersymmetric models with
$R$-parity conservation and with the lightest neutralino
$\chi$ as the lightest supersymmetric particle (LSP). In particular, 
the configuration in which the mass difference
$\Delta m$ between the charginos and the LSP is less than
5\,GeV/$c^2$ is studied. 

The small-$\Delta m$ configuration is possible in the
MSSM, the minimal supersymmetric extension of the Standard Model
\cite{hab}, although it requires unusual assumptions to be made
for the gaugino mass terms, $M_1$ and $M_2$. Indeed, with the
usual gaugino-mass unification relation, charginos and neutralinos
are nearly mass degenerate only in the deep higgsino region, {\it
i.e.}, for very large $M_2$ values. However, it happens for more
natural values of the MSSM parameters if the gaugino-mass
unification assumption is relaxed. It even becomes common in
models with anomaly-mediated supersymmetry breaking \cite{giu}, in
which $M_1$ is naturally much larger than $M_2$.

Charginos generally decay into the stable LSP and a pair
of fermions ($\chi {\rm q\bar q}^\prime$ or
$\chi \ell\nu_\ell$). In the small-$\Delta m$ configuration, 
three different final state topologies may occur 
according to the value of $\Delta m$.
\begin{enumerate}
\item[1.] The mass difference, and therefore the phase space available for
the chargino decay, is so small that charginos are long-lived. 
A search for heavy stable charged particles~\cite{stable,stable_npy}
is efficient for this topology.
\item[2.] The mass difference is sufficiently large for the chargino decay
products to trigger the data acquisition. In this case, the final
state is characterized by the large amount of missing energy
carried by the invisible LSPs, as searched for in Refs.~\cite{emiss,emiss_npy}.
\item[3.] The mass difference is in between, {\it i.e.}, large enough
for the charginos to decay before leaving the detector, but too
small for the decay products to trigger the data acquisition.
\end{enumerate}
The last, intermediate, situation is analysed in this letter. As
suggested in Ref.~\cite{drees}, a search for radiative chargino
pair production, ${\rm e}^+{\rm e}^- \to \gamma\, \chi^+\chi^-$,
was performed, where the initial state radiation (ISR) photon is
 emitted at sufficiently large angle with respect to the
incoming beam to be detected, and with sufficiently large energy
to activate the trigger system. The relevant topology is therefore
an energetic, isolated photon accompanied by a few low-momentum
particles from the chargino decays and large missing energy
carried by the two neutralinos. This topology has also been
searched for by other LEP collaborations~\cite{DELPHI,L3} with
lower energy data.

This letter is organized as follows. In Section~2, the ALEPH
detector elements directly related to the ISR photon search are 
described, followed by a summary of
the signal and background simulation in Section~3. The event
selection developed for the ISR photon final state is explained in
Section~4, the results are presented in Section~5, and their
interpretation in the MSSM, combined with those of the other two
topological searches mentioned above, is given in Section~6.

\begin{table}[htbp]
\begin{center}
\def\baselinestretch{1.2} 
\caption{\footnotesize Integrated luminosities collected
between 1997 and 2000, average centre-of-mass energies, and data
samples used by the analyses mentioned in the text.
\label{tab:lumi}} \vspace{3mm}
\def\baselinestretch{1.}
\fontsize{10.4745}{12pt}\selectfont
\begin{tabular}{||l|c|c|c|c|c|c|c|c|c|c||} \hline
Year & 1997 & 1998 & \multicolumn{4}{|c|}{1999} &
\multicolumn{4}{|c||}{2000}
\\ \hline\hline
$\sqrt{s}$ (GeV) & 182.7 & 188.6 & 191.6 & 195.5
                 & 199.5 & 201.6 & 203.2 & 205.0 & 206.5 & 208.0 \\
\hline ${\cal L}$ (${\rm pb}^{-1}$)
                 &  56.8 & 174.2 & 28.9  &  79.8
                 &  86.2 &  42.0 & 11.6 & 71.6  & 126.3 & 7.3 \\
\hline\hline Stable charginos &
\multicolumn{2}{|c|}{Ref.~\cite{stable}} &
\multicolumn{8}{|c||}{Ref.~\cite{stable_npy}}\\ \hline Missing energy &
\multicolumn{6}{|c|}{Ref.~\cite{emiss}} &
\multicolumn{4}{|c||}{Ref.~\cite{emiss_npy}} \\ \hline ISR photons &
\multicolumn{1}{|c|}{---} & \multicolumn{9}{|c||}{This analysis}
\\
\hline\hline
\end{tabular}
\end{center}
\end{table}

\section{The ALEPH detector}

A complete and detailed description of the ALEPH detector and its
performance, as well as of the standard reconstruction and
analysis algorithms, can be found in Refs.~\cite{det,perf}. Only
those items relevant for the final state under study in this
letter (an ISR photon accompanied by a few low-momentum particles
and substantial missing energy) are described below.

The hermetic electromagnetic calorimeter, a
22-radiation-length-thick sandwich of lead planes and proportional
wire chambers with fine readout segmentation, consists of 36
modules, twelve in the barrel and twelve in each endcap. It is
used to identify photons and to measure their energies, with a 
relative resolution of $0.18/\sqrt{E} + 0.009$ ($E$~in~GeV), 
and their positions down to $13^\circ$ from the beam axis. 
Unconverted photons are reconstructed as localized energy 
deposits (clusters) within
groups of neighbouring cells in at least two of the three segments
in depth of the calorimeter. The longitudinal and transverse
distributions of these deposits are required to be consistent with
those of an electromagnetic shower. The \emph{impact parameter}
with respect to the interaction point is calculated for a given
cluster from the barycentres of its energy deposits in each
segment. The \emph{compactness} of a cluster is calculated by
taking an energy-weighted average of the angle subtended at the
interaction point between the barycentre of the whole cluster and
the barycentre of its deposit in each segment.

Only those photons with a polar angle such that $\vert
\cos\theta_\gamma \vert< 0.95$, with an energy in excess of
1\,GeV, with an interaction time relative to a beam crossing
smaller than 100\,ns, with an impact parameter smaller than
$80$\,cm and with a compactness smaller than 1$^\circ$ are
considered as ISR photon candidates in this letter. The energy
deposits associated with photons from bremsstrahlung of high energy
particles in cosmic ray events are rejected by the requirements
on interaction time, impact parameter and compactness. 
Finally, ISR photon candidates are required to
be isolated from any reconstructed charged particle by more than
$30^\circ$.

Charged particles are detected in the central tracking system,
which consists of a silicon vertex detector, a cylindrical
multiwire inner drift chamber (ITC) and a large time projection
chamber (TPC). It is immersed in a 1.5\,T axial magnetic field
provided by a superconducting solenoidal coil surrounding the
electromagnetic calorimeter. Photons traversing the tracking
material may convert into electron-positron pairs (in 6\% of the cases 
at normal incidence) and can therefore not be identified
as depicted above. To identify these photons, pairs are formed of
two oppositely charged particle tracks reconstructed with at least
four hits in the TPC and extrapolated to a common vertex; at this
position the momenta are computed and the invariant mass of the
pair is determined assuming electron masses. Photon conversions
are identified if this invariant mass is smaller than $100 \, \mem
$. The same acceptance, energy and isolation criteria are applied
as to unconverted photons to define an ISR photon candidate.

Other charged particles in the event are called {\it good tracks}
if they are reconstructed with at least four hits in either the
ITC or the TPC, and if they
originate from within a cylinder of length 10\,cm (for tracks
reconstructed with at least four TPC hits) and radius 2\,cm (4\,cm
for tracks reconstructed with no TPC hits) coaxial with the beam
and centred at the nominal interaction point.

Global event quantities such as total energy (and therefore
missing energy) are determined from an energy-flow algorithm which
combines the above measurements supplemented by those of the iron
return yoke instrumented as a hadron calorimeter, the surrounding
two double layers of muon chambers, and the luminosity monitors, 
which extend the calorimetric coverage down to 34\,mrad. This
algorithm provides, in addition, a list of {\it energy-flow
particles}, classified as photons, electrons, muons, and charged and
neutral hadrons, which are the basic objects used in the selection
presented in this letter.

Finally, use is made of the hadron calorimeter to improve
the energy resolution of photons pointing to cracks between
modules of the electromagnetic calorimeter. To do so, the energy
of any neutral energy-flow particle is added to that of an ISR
photon candidate if it is within $11.5^\circ$ of the
photon direction.

The trigger condition relevant for this analysis consists of 
a deposit of at least 1\,GeV (2.3\,GeV) in any single
 module of the barrel (the endcaps) of the electromagnetic calorimeter. 
This trigger is fully efficient for ISR photons within the acceptance 
of the present search.

\section{Event simulation}
Background events expected from Standard Model processes were
generated with statistics corresponding to at least 20 times the 
integrated luminosity of the data. 
The event generators used for the present analysis are
listed in Table~\ref{bkg}.

Signal events were simulated with the program {\tt
SUSYGEN}~\cite{montecarlo}. Charginos were produced with masses
ranging from 45\,GeV/$c^2$ up to the kinematic limit, for mass
differences with the LSP between $m_{\rm e}$ and 5\,GeV/$c^2$ and
with proper decay length $\lambda=c \tau$ up to 80\,cm. Hadronic
decays were modelled as suggested in Ref.~\cite{cheen}. For
$\Delta m < 2 \, \gem$, the model of Ref.~\cite{kuhn} was used for 
the spectral functions, with
parameters tuned to agree with the measured hadronic $\tau$
decays~\cite{tau}. In this case, chargino decays
to ${\rm e}\nu_{\rm e}\chi $, $\mu \nu_\mu\chi$ and
$n \pi \, \chi $ ($n=1, 2, 3$) were simulated.
For larger $\Delta m$ values, the Lund
fragmentation scheme was applied, as implemented in {\tt{SUSYGEN}}.
Initial state radiation was simulated according to the treatment
described in Ref.~\cite{ISR}.

\def\baselinestretch{1.}
\begin{table}[htbp]
\caption{\footnotesize{Standard Model background processes and
the generators used to simulate them in the present analysis.}} \label{bkg}
\begin{center}
\begin{tabular}{||l|l||}  \hline
 Standard Model processes & Generators \\
\hline\hline $ \mathrm{e}^+ \mathrm{e}^- \rightarrow
\gamma (\gamma) \nu \bar{\nu} $ & {\tt{KORALZ}} \cite{KORALZ} \\
\hline $
\mathrm{e}^+ \mathrm{e}^- \rightarrow \mathrm{e}^+ \mathrm{e}^-$ & {\tt{BHWIDE}} \cite{BHWIDE} \\
\hline
$ \mathrm{e}^+ \mathrm{e}^- \rightarrow \mu^+ \mu^- , \, \tau^+ \tau^- $ & {\tt{KORALZ}} \\
\hline $ \mathrm{e}^+ \mathrm{e}^- \rightarrow \mathrm{q} \bar{\mathrm{q}}  $ & {\tt{KORALZ}}  \\
\hline
$\mathrm{e}^+ \mathrm{e}^- \rightarrow \mathrm{W}^+ \mathrm{W}^- $ &  {\tt{KORALW}} \cite{KORALW} \\
\hline $ \mathrm{e}^+ \mathrm{e}^- \rightarrow \mathrm{W}
\mathrm{e} \nu , \, \mathrm{Z} \mathrm{Z} , \, \mathrm{Z}
\mathrm{e}^+
\mathrm{e}^- , \, \mathrm{Z} \nu \bar{\nu} $ &  {\tt{PYTHIA}} \cite{PYTHIA} \\
\hline $ \gamma \gamma \rightarrow \mathrm{e}^+ \mathrm{e}^- ,\,
\mu^+ \mu^- , \, \tau^+
\tau^- $ &{\tt{PHOT02}} \cite{PHOT02} \\
\hline $ \gamma \gamma \rightarrow \mathrm{q} \bar{\mathrm{q}} $ & {\tt{PHOT02}}  \\
\hline \hline
\end{tabular}
\end{center}
\end{table}
\def\baselinestretch{1.}

\section{Event selection}

Events are selected if they contain at least one ISR photon
candidate (a calorimeter photon or a photon conversion)
reconstructed with transverse momentum $p_T^\gamma$ with respect 
to the beam axis greater than $0.035 \sqrt{s}$. This requirement
effectively rejects the background from $\gamma\gamma$ interactions, 
further reduced by requiring that no energy be detected within 
$14^\circ$ of the beam axis. This last cut is also useful to suppress
background from Bhabha scattering.

At least two charged particle tracks with a minimum of four ITC or four 
TPC hits are required, and at most ten good tracks. The latter cut efficiently 
reduces the backgrounds from hadronic two- and four-fermion processes.
In events from ${\rm e}^+{\rm e}^- \to\gamma\gamma\nu\bar\nu$, a photon 
may convert at the ITC/TPC boundary, but fail to be identified by the 
criteria of Section~2 while still being reconstructed as two charged 
particles. This background is rejected by requiring that at least one 
track be reconstructed with at least one ITC hit.

Advantage is taken of the characteristic kinematic features of radiative 
chargino pair production. The ISR photon energy is required to be smaller 
than the maximum energy allowed in the process ${\rm e}^+{\rm e}^- \to
\gamma \chi^+\chi^-$, and the leading charged particle momentum is limited 
to the maximum allowed chargino momentum. The mass recoiling against the 
ISR photon must be larger than 100\,GeV/$c^2$. Sliding upper cuts are applied 
to the missing transverse momentum and to the energy of the chargino
decay products as a function of the hypothetical chargino mass and
mass difference with the LSP; these cuts were determined from the
simulation of promptly decaying charginos.

\section{Results and systematic uncertainties}
\begin{figure}[t]
\setlength{\unitlength}{1.0mm}
\begin{center}
\begin{picture}(160,200)(0,0)
\put(0,130){\mbox{\psfig{figure=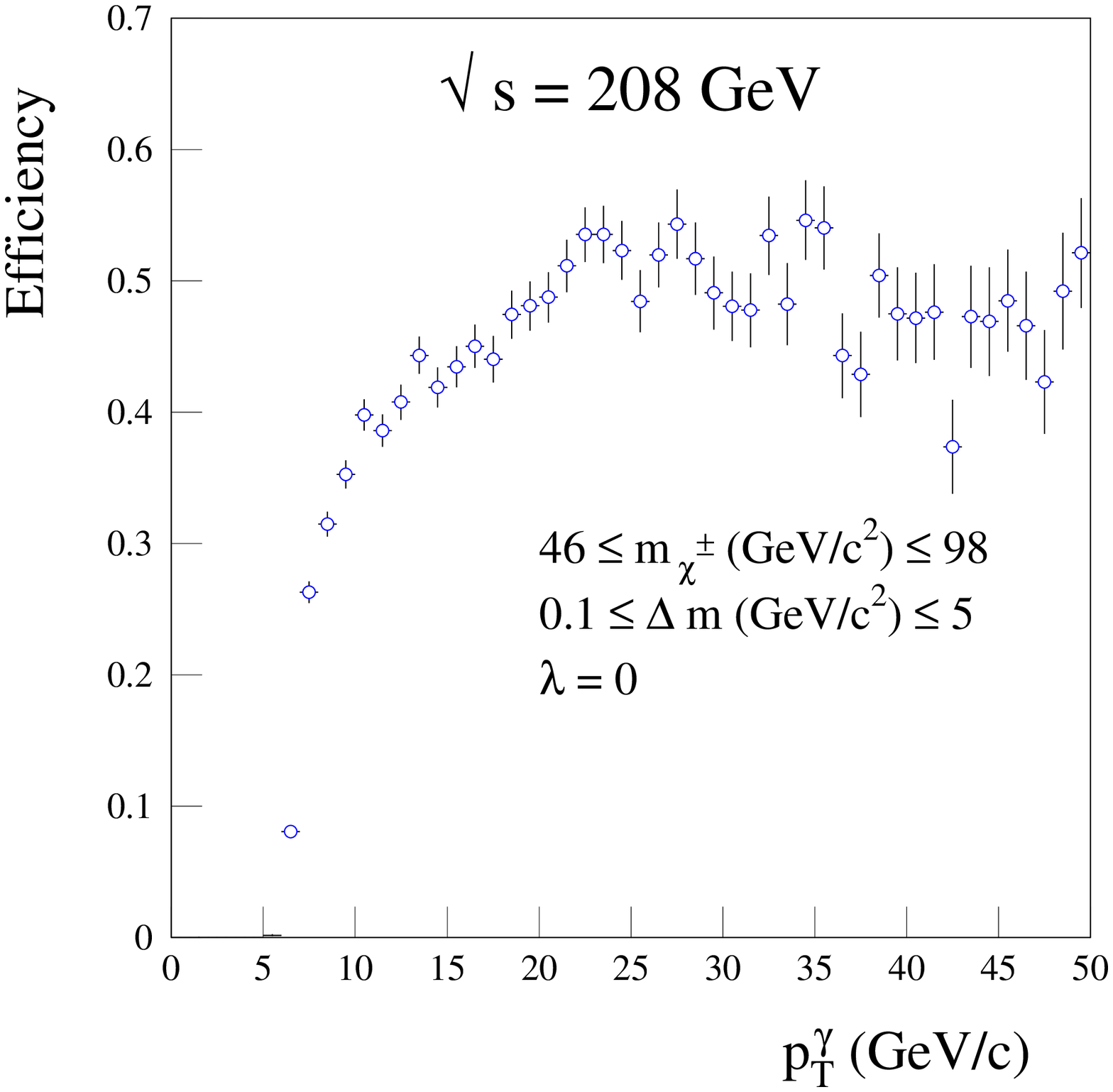,width=8.5cm,height=8.2cm}}}
\put(82,130){\mbox{\psfig{figure=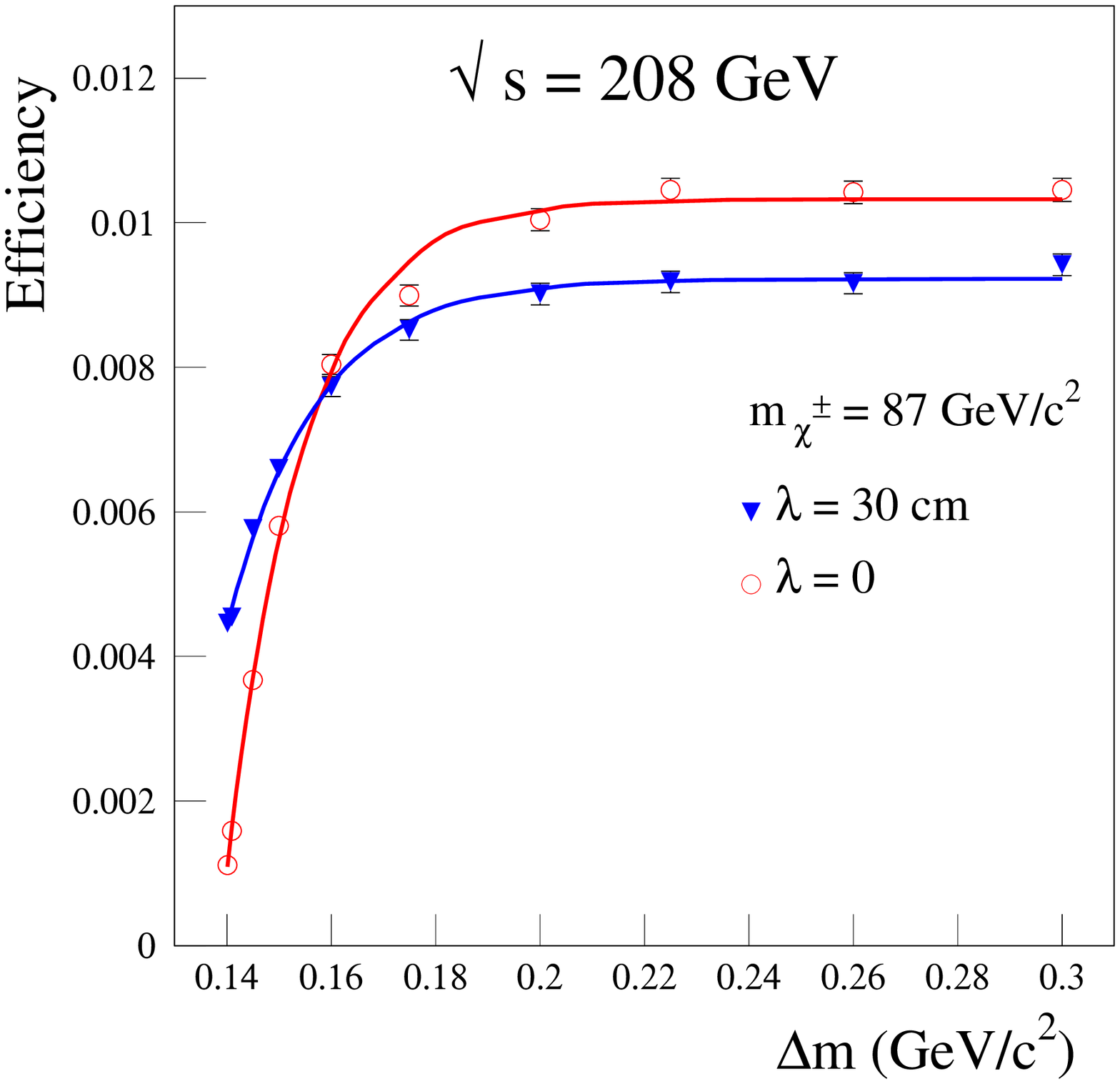,width=8.5cm,height=8.2cm}}}
\put(0,40){\mbox{\psfig{figure=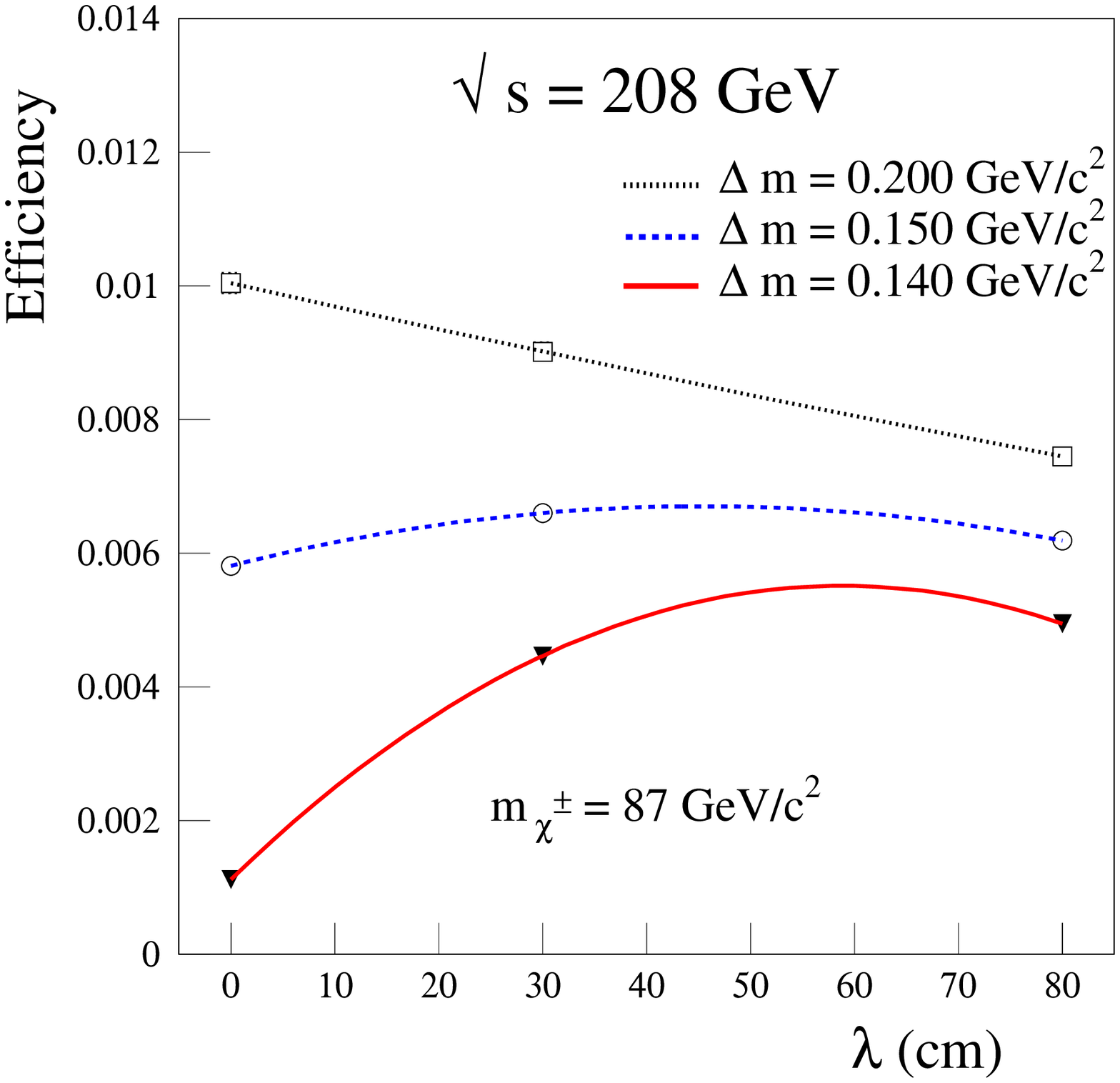,width=8.5cm,height=8.2cm}}}
\put(82,40){\mbox{\psfig{figure=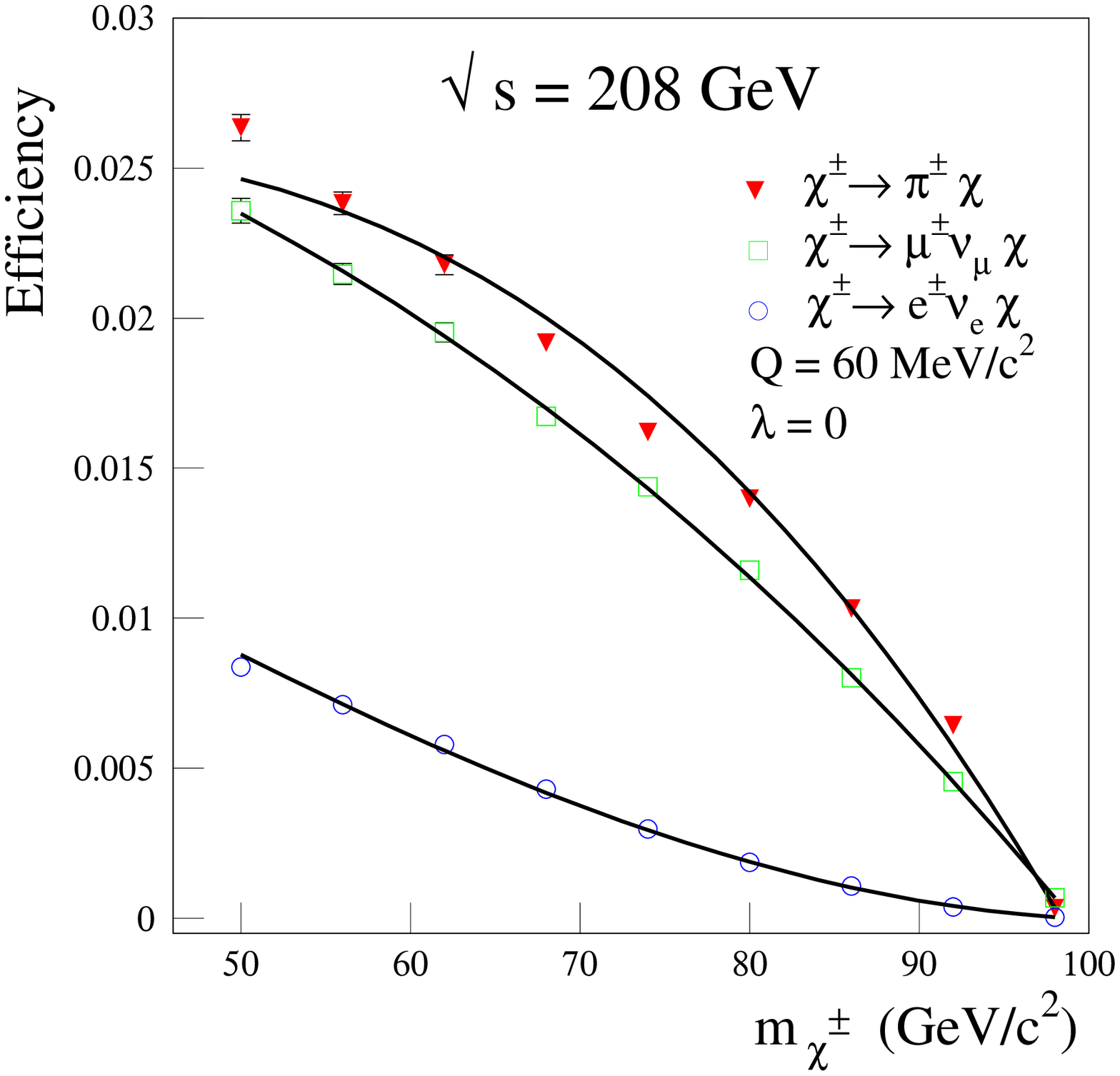,width=8.5cm,height=8.2cm}}}
\put(71,198){a)} \put(153,198){b)}
\put(71,108){c)}\put(153,108){d)}
\end{picture}
\end{center}
\vspace{-3.5cm}
\def\baselinestretch{1.2}
\caption{\footnotesize{a) Selection efficiency as a function of the 
generated transverse momentum of the
ISR photon produced in association with charginos decaying
promptly with masses between 46 and $98~\gem$, $\Delta m$ between
0.1 and 5 $\gem$ at $\sqrt{s} = 208~\gee$. b) Signal efficiency as
a function of $\Delta m $ at $\sqrt s = 208 \, \gee$,
$m_{\charg}=87 \, \gem$ and $\lambda=0, \, 30 \, \mathrm{cm}$ for
the single-pion final state. c) Signal efficiency as a function of
$\lambda $ at $\sqrt s = 208 \, \gee$, $m_{\charg}=87 \, \gem$ and
$\Delta m = 0.14, \, 0.15, \, 0.2 \, \gem$. d) Signal efficiency
as a function of the chargino mass at $\sqrt s = 208~\gee$ for
${Q}= 60 \, \mem $, $\lambda=0$, for the single-pion and leptonic final
states.}} \label{Effi}
\def\baselinestretch{1.}
\end{figure}

The selection efficiency is shown in Fig.~\ref{Effi}a as a
function of the generated $p_{\rm T}^\gamma $. 
For $p_{\rm T}^\gamma$ in excess of 15\,GeV/$c$, the
selection efficiency is above 45\%. The photon identification
criteria described in Section~2 (acceptance, reconstruction,
isolation, timing and pointing) are responsible for an efficiency
loss of 40\%, and the remaining loss is due to the event selection of
Section~4. Because the $p_{\rm T}^\gamma$
spectrum is peaked at small values, the total
selection efficiency is at most 3\%.

The dependence of the selection efficiency on $p_{\rm T}^\gamma$
is qualitatively identical for all chargino masses studied, but
the absolute efficiency level depends also on the chargino decay
length and on the available phase space $Q = \Delta m - \sum_i
m_i$, where the sum runs over the chargino decay products.

For instance, the efficiency for promptly-decaying charginos with
mass 87\,GeV/$c^2$ is shown in Fig.~\ref{Effi}b as a function of
$\Delta m$ for the single pion final state, and is compared to
that for charginos with a proper decay length $\lambda$ of 30\,cm.
The higher efficiency for larger decay lengths for very low ${Q}$
is due to the detection of the chargino tracks. 
For the same $\lambda$, at larger $\Delta m$, the chargino
detection efficiency is lower with respect to the case $\lambda=0$
because of the cuts on the reconstructed energy in addition to the
photon and on the total visible momentum of the event. The
dependence on the proper decay length is displayed in
Fig.~\ref{Effi}c, for $\Delta m =$ 140, 150 and 200\,MeV/$c^2$.

The dependence of the absolute efficiency on the chargino mass 
(through its effect on the  $p_{\rm T}^\gamma$ spectrum) is shown in
Fig.~\ref{Effi}d, for three different final states (electron, muon
and single pion), for fixed $Q$ values and $\lambda=0$. At the
same ${Q}$, the decay into an electron is reconstructed less efficiently
than the decay into a muon due to the mass of the final state lepton, and
the efficiency for the three-body decays is smaller than for
the two-body decay to a single pion.

The photon spectrum is also strongly dependent on the field
content of the chargino which is different in various regions of
the MSSM parameter space. Initial state radiation 
is enhanced in the gaugino region
due to the relative contribution of the $s$-channel Z exchange.
All the plots shown in Fig.~\ref{Effi} were derived in the gaugino
region.

\def\baselinestretch{1.}
The efficiency includes a $(-5\pm 1)$\% correction for the loss 
due to the cut on the energy measured within
$14^\circ$ of the beam axis, caused by beam-related background in
the forward calorimeters. This correction is determined with
events triggered at random beam crossings.

A systematic uncertainty of 0.6\% on the efficiency, related to
the algorithm of photon reconstruction, is taken into account as
described in Ref.~\cite{gary}. The uncertainty on the number of
converted photons is estimated to change the total selection
efficiency by up to 0.3\%.

Systematic uncertainties from the ISR photon simulation are
assessed by comparing the ISR photon transverse momentum spectrum as
predicted by the \texttt{SUSYGEN} and {\tt{KORALZ}} programs. The
distributions obtained from {\tt{KORALZ}} for the mass recoiling against 
the photon and for the
polar angle of the photon in single photon events are in
agreement with data~\cite{gary}. The systematic uncertainty on the
efficiency of the present selection is estimated by integrating
the possible discrepancy of the $p_T^{\gamma}$ distributions between 
the two generators over
the allowed range, which depends on the chargino mass. The systematic
uncertainty obtained is at most 10\%.

The systematic error on the efficiency due to the limited
statistics of the simulated samples is about 3\%.
The total systematic uncertainty on the selection efficiency is 10\%,
obtained by adding in quadrature the individual contributions.

The numbers of events observed are in agreement with those
expected from Standard Model background sources, dominated by
${\rm e}^+{\rm e}^- \to \tau^+ \tau^-$, $\gamma \gamma \to \tau^+
\tau^-$ and ${\rm e}^+{\rm e}^- \to \gamma (\gamma) \nu \bar{\nu}
$, as shown in Table~\ref{number} for several chargino masses and
mass differences. These numbers are displayed in Fig.~\ref{obsexp}
in the $(m_{\chi^\pm}, \Delta m)$ plane. A candidate event that
contributes up to $m_{\charg} = 84~\gem$, independent of $\Delta m$, is
shown in Fig.~\ref{evento}.

\def\baselinestretch{1.}
\begin{center}
\begin{table}[htbp]
\caption{\footnotesize{Numbers of events observed and expected from 
Standard Model background sources for $m_{\charg} > 50,\ 65,\ 85~\gem$ and 
$\Delta m < 2,\ 1,\ 0.3~\gem$, respectively. The main
contributions to the expected background are also reported.}}
\label{number}
\vspace{0.35cm}
\fontsize{7.9}{12pt}\selectfont
\begin{tabular}{||c|c||c|c|c|c|c|c|c||}
 \hline
 \multicolumn{2}{||c||}{$\sqrt s = 189$--$209 \, \mathrm{GeV} $ } &
Data & \multicolumn{6}{c||}{Background}  \\
\hline \hline $m_{\charg}$ & $\Delta m$ &
$\mathrm{N}_{\mathrm{obs}}$ & $\epem \rightarrow \gamma (\gamma)
\nu \bar{\nu} $ & $\gamma \gamma \to \tau^+ \tau^-$ & $ \gamma
\gamma \to \epem $ & $ \mathrm{e}^+ \mathrm{e}^- \rightarrow
\tau^+ \tau^-$ &
$\rm{four \, \, fermions}$ & $\mathrm{N}_{\mathrm{exp}}$  \\
\hline \hline $> 50$ & $< 2   $ & 13 & 1.2 & 2.0 & 0.3 & 3.8 & 1.3
&  {9.0}  \\ \hline $> 65$ & $< 1   $ &  5 & 0.7 & 1.8 & 0.3 & 1.7
& 0.6 &  {5.4}  \\ \hline $> 85$ & $< 0.3 $ &  1 & 0.6 & 1.4 &   0
& 0.5 &  0  &  {2.9}  \\ \hline\end{tabular}
\end{table}
\end{center}

\begin{figure}[t]
\setlength{\unitlength}{1.0mm}
\begin{center}
\begin{picture}(160,200)(0,0)
\put(24,130){\mbox{\psfig{figure=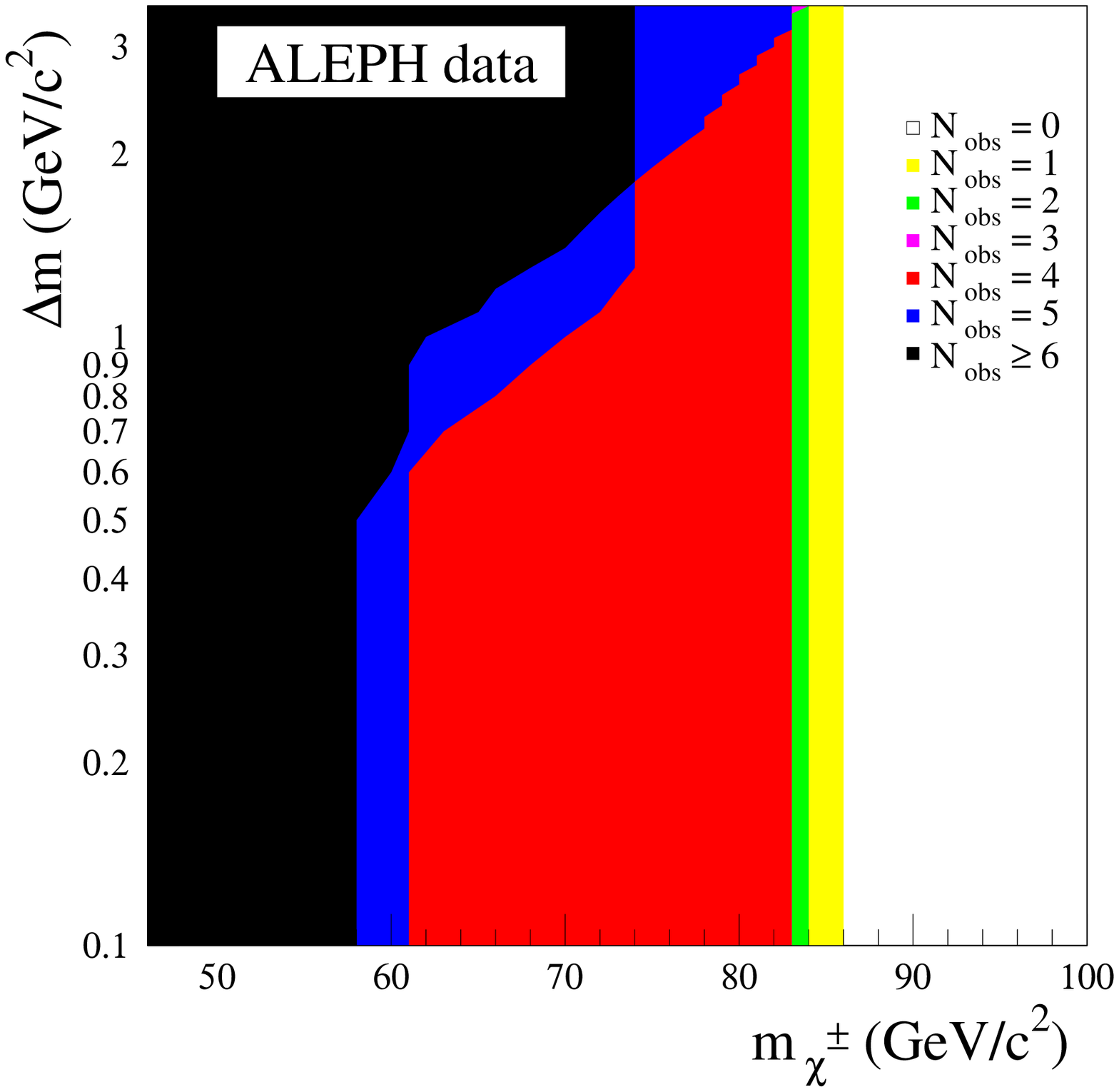,width=11cm,height=10cm}}}
\put(24,30){\mbox{\psfig{figure=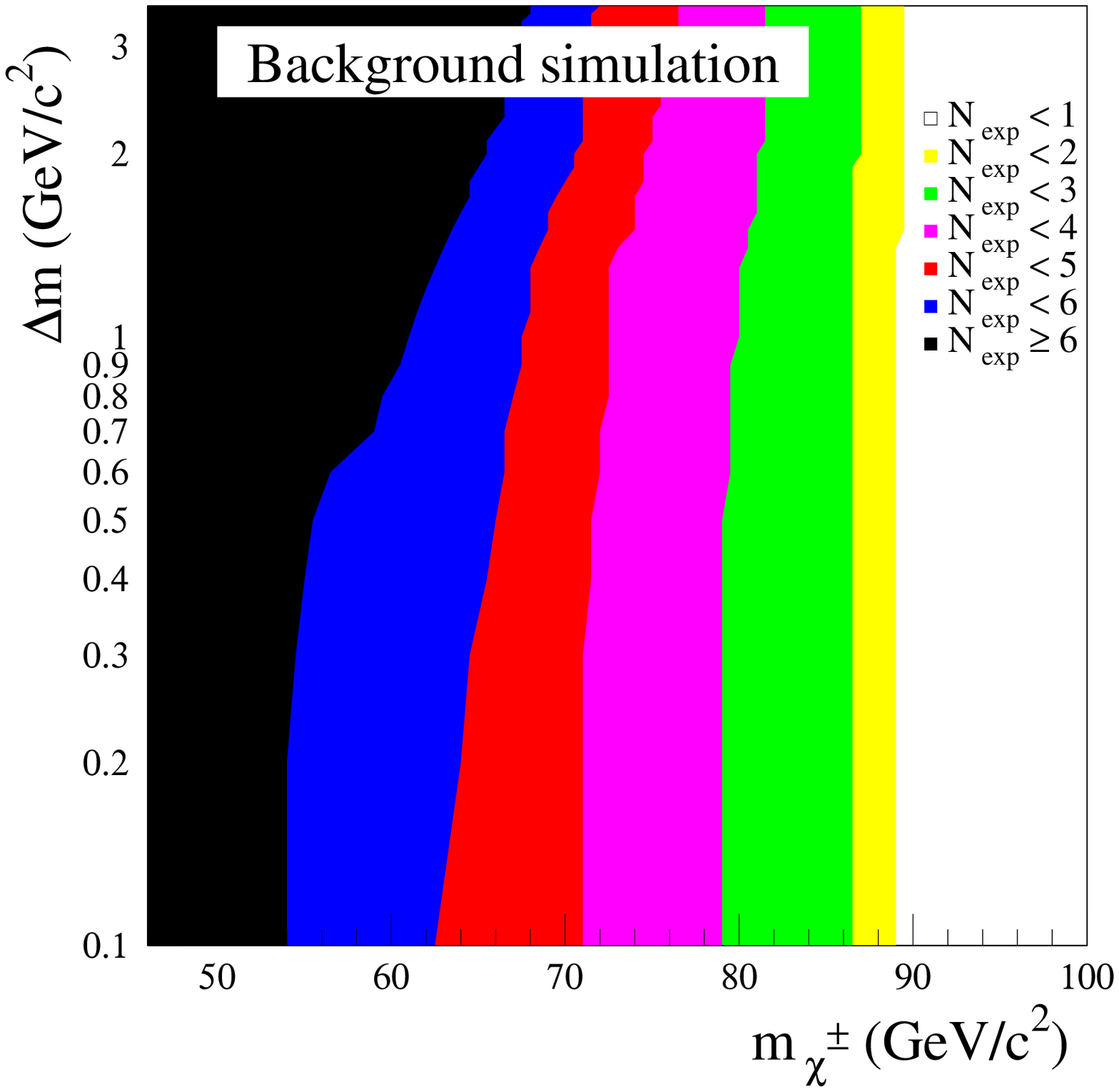,width=11cm,height=10cm}}}
\put(118,215){a)} \put(118,115){b)}
\end{picture}
\end{center}
\vspace{-3.5cm}
\def\baselinestretch{1.2}
\caption{\footnotesize{Number of events a) observed and b)
expected from standard background sources in the
($m_{\charg}, \Delta m$) plane. The number of events decreases
for large chargino masses or low $\Delta m$ due to the tightening
of the cuts on the transverse missing momentum and on the total
energy not associated with the reconstructed photons. No events compatible 
with $ m_{\charg} > 86 \, \gem $ are observed in the data.}}
\label{obsexp}
\def\baselinestretch{1.}
\end{figure}

\section{Interpretation in the MSSM}

In order to provide complete coverage of all possible chargino
decay lengths, the results of the present analysis, hereafter
called {\it the ISR analysis}, were combined with those of the
search for heavy stable charged
particles~\cite{stable,stable_npy}, needed for very small $\Delta
m$ values, and the standard missing-energy
search~\cite{emiss,emiss_npy}, efficient for $\Delta m \ge 3$\,GeV/$c^2$. 
No background subtraction was performed and
systematic uncertainties were taken into account according to
Ref.~\cite{syse}.

In the MSSM, and in the absence of the assumption of gaugino-mass
unification, four parameters are to be considered in the gaugino
sector: the soft-breaking gaugino mass terms, $M_1$ and $M_2$; the
Higgs mixing term, $\mu$; and the ratio of the two Higgs-doublet
vacuum expectation values, $\tan\beta$. To determine 95\%
confidence level (C.L.) excluded  regions in the
$(m_{\chi^\pm},\Delta m)$ plane, a broad scan of these parameters
was performed: $1 < \tan\beta < 300$, $\vert \mu \vert <
1$\,TeV/$c^2$, $M_2 < 250$\,TeV/$c^2$, and $M_1$ chosen so as to
cover the range $\Delta m < 5$\,GeV/$c^2$.

The regime in which all sfermion masses are very large is
considered first. In this case, the chargino production cross section
depends only on the four above-mentioned parameters, and charginos
generically decay {\rm via} $\chi^\pm \to \chi {\rm
W}^{\pm\ast}$.

\begin{figure}[t]
\setlength{\unitlength}{1.0mm}
\begin{center}
\begin{picture}(160,200)(0,0)
\put(24.,130){\mbox{\psfig{figure=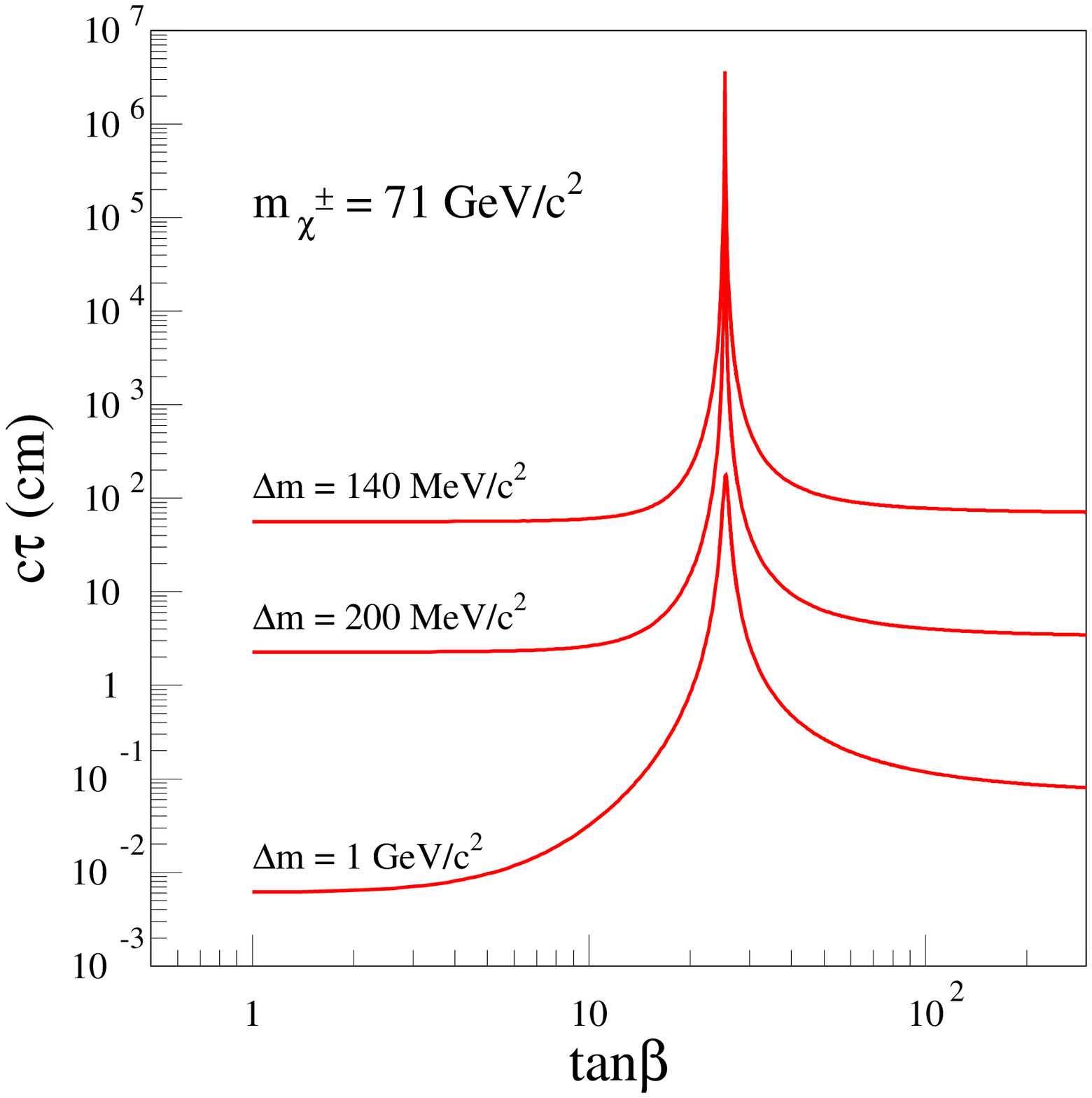,width=11cm,height=10cm}}}
\put(24,30){\mbox{\psfig{figure=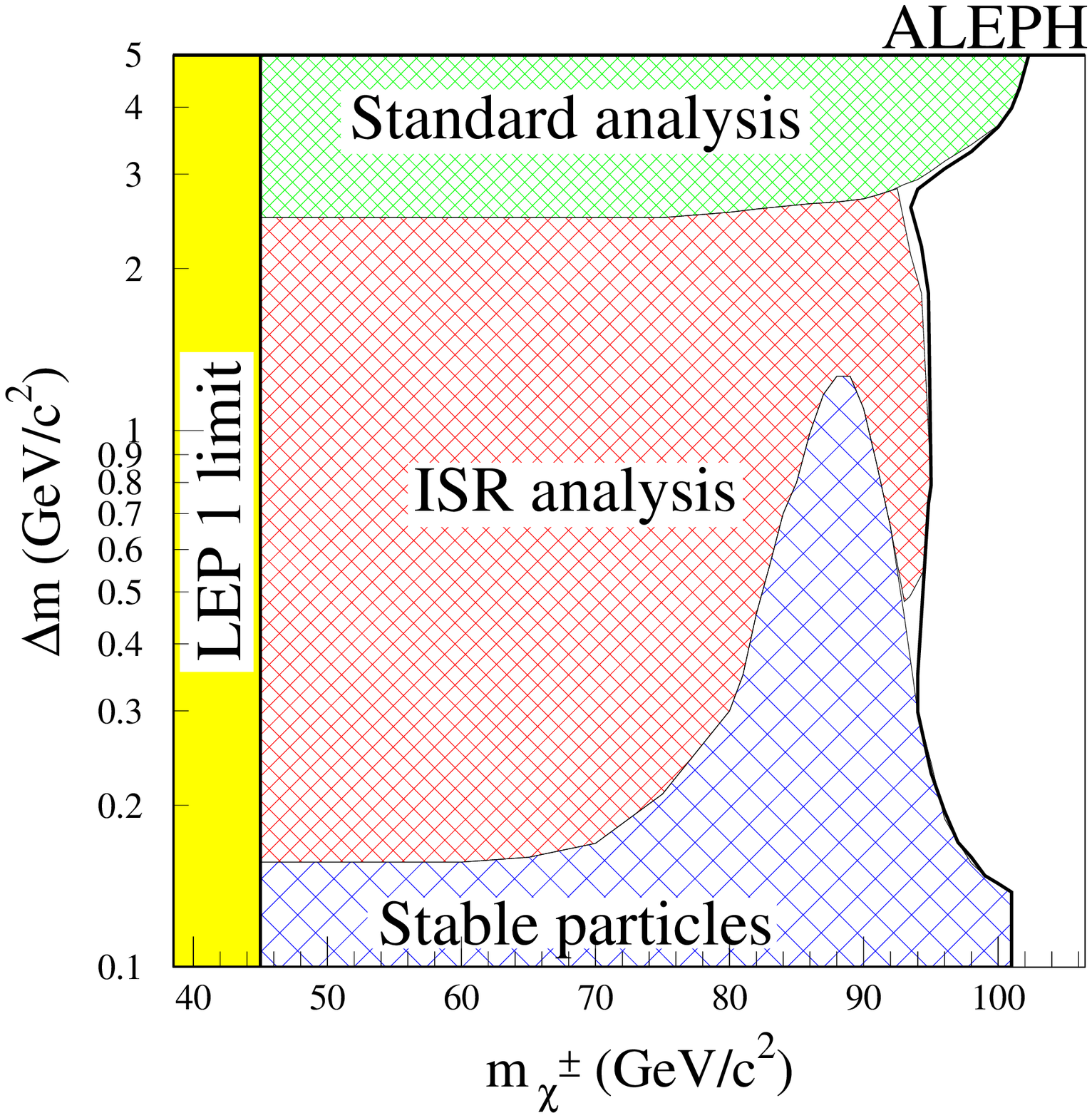,width=11cm,height=10cm}}}
\put(118,215){a)} \put(118,114){b)}
\end{picture}
\end{center}
\vspace{-3.5cm}
\def\baselinestretch{1.2}
\caption{\footnotesize{a) Chargino proper decay length as a
function of $\tan \beta$ for $m_{\charg} = 71 \, \gem$, 
for $m_0 = 500 \, \gem$ and for $\Delta m=140, \ 200, \ 1000~\mem$. 
b) Exclusion
region in the ($m_{\charg} , \Delta m$) plane at 95\% C.L., in the
large scalar mass scenario ($m_0=500 \, \gem $) and in the gaugino
region, for $\tan \beta = 21 $. The top area is excluded by the
standard missing energy chargino selection~\cite{emiss,emiss_npy} while the
bottom area is excluded by the stable particle
search~\cite{stable,stable_npy}. The ISR analysis covers the intermediate
area. The region excluded by the combination of the three analyses
is bounded by the bold curve. The region excluded at LEP\,1 is
also shown.}} \label{lifetime}
\def\baselinestretch{1.}
\end{figure}

\begin{figure}[p]
\begin{center}
\includegraphics*[width=15.cm, height=14.cm, bb=14 156 545 674]{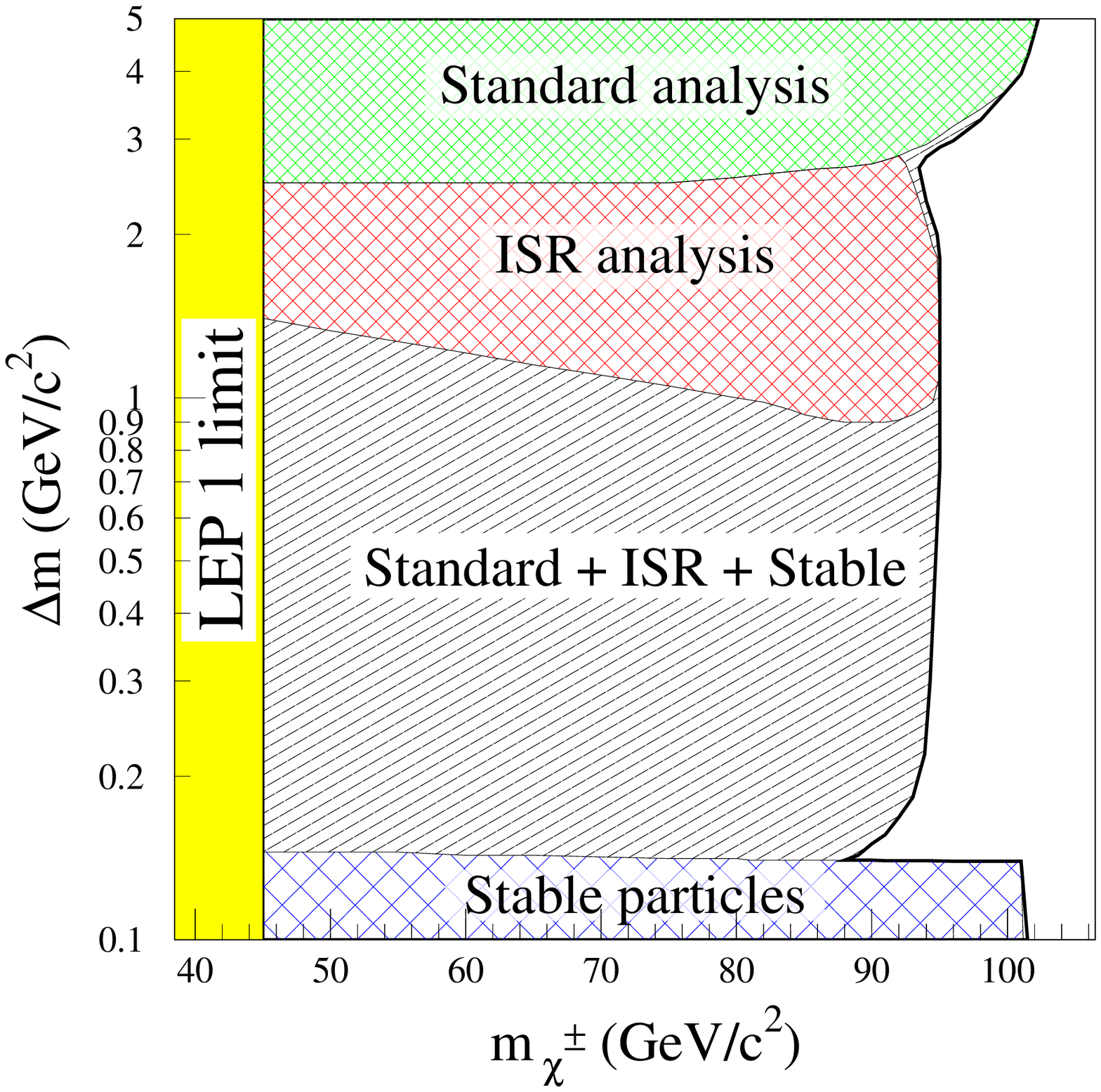}
\def\baselinestretch{1.2}
\caption{\footnotesize{Excluded region in the ($m_{\charg},\Delta m$) plane 
at 95\% C.L., in the large scalar mass scenario
($m_0=500 \, \gem$) and in the gaugino region, independently of
$\tan \beta$. The top area is excluded by the standard missing
energy chargino selection~\cite{emiss,emiss_npy} while the bottom area is
excluded by the stable particle search~\cite{stable,stable_npy}. The ISR
analysis provides the exclusion for $\Delta m \sim 1$~to~$2.5 \,
\gem $. The remaining $\Delta m$ region is excluded by the
combination of the ISR analysis and of the stable particle search.
The region excluded by the combination of the three analyses is
bounded by the bold curve. The region excluded at LEP\,1 is
also shown.}} \label{ISRgaug}
\end{center}
\end{figure}

In the gaugino region ($M_2 \ll \vert\mu\vert$), it is possible to
fine-tune $\tan\beta$ to make the $\chi^\pm \chi {\rm W^\pm}$
coupling vanish, in which case the chargino decay length increases
substantially, even for large $\Delta m$ values. As a result, the
ISR analysis is efficient independently of $\tan\beta$ only over a
limited range of $\Delta m$, the extent of which depends on the
$\chi^+ \tilde{\rm f} \bar{\rm f}^\prime$ couplings. This effect was
quantitatively investigated under the assumption of a universal
sfermion mass term $m_0$. The dependence on $\tan \beta$ of the
chargino decay length is shown for $m_{\charg}=71 \, \gem$ 
in Fig.~\ref{lifetime}a, for $m_0 = 500 \, \gem$ and for various 
$\Delta m$ values. The exclusion domain in
the ($m_{\charg}, \Delta m$) plane derived at $m_0=500 \, \gem$
and fixed $\tan \beta = 21$ (which corresponds to the maximum
decay length for a chargino mass of $88 \, \gem$) is shown in
Fig.~\ref{lifetime}b; the ISR analysis covers the gap between the
standard missing energy search and the stable particle analysis.
Results independent of $\tan \beta$ are also shown in
Fig.~\ref{ISRgaug} for $m_0=500\, \gem$. The standard missing
energy search excludes $\Delta m$ values larger than $\sim
2.3$\,GeV/$c^2$, and the ISR analysis $\Delta m$ values down to
1~to~1.5\,GeV/$c^2$. The search for heavy stable charged particles
is fully efficient in a parameter-independent way only for $\Delta
m < m_\pi$, but a combination with the ISR analysis allows the
exclusion of the intermediate $\Delta m$ range to be achieved,
independently of $\tan\beta$. In the end, a 95\%~C.L. lower limit
on the chargino mass of 88\,GeV/$c^2$ is set in the gaugino region
for heavy sfermions ({\it i.e.,} for sfermion masses, or $m_0$,
larger than a few hundred GeV/$c^2$).

\begin{figure}[p]
\begin{center}
\includegraphics*[width=15.cm, height=14.cm, bb=14 156 545 674]{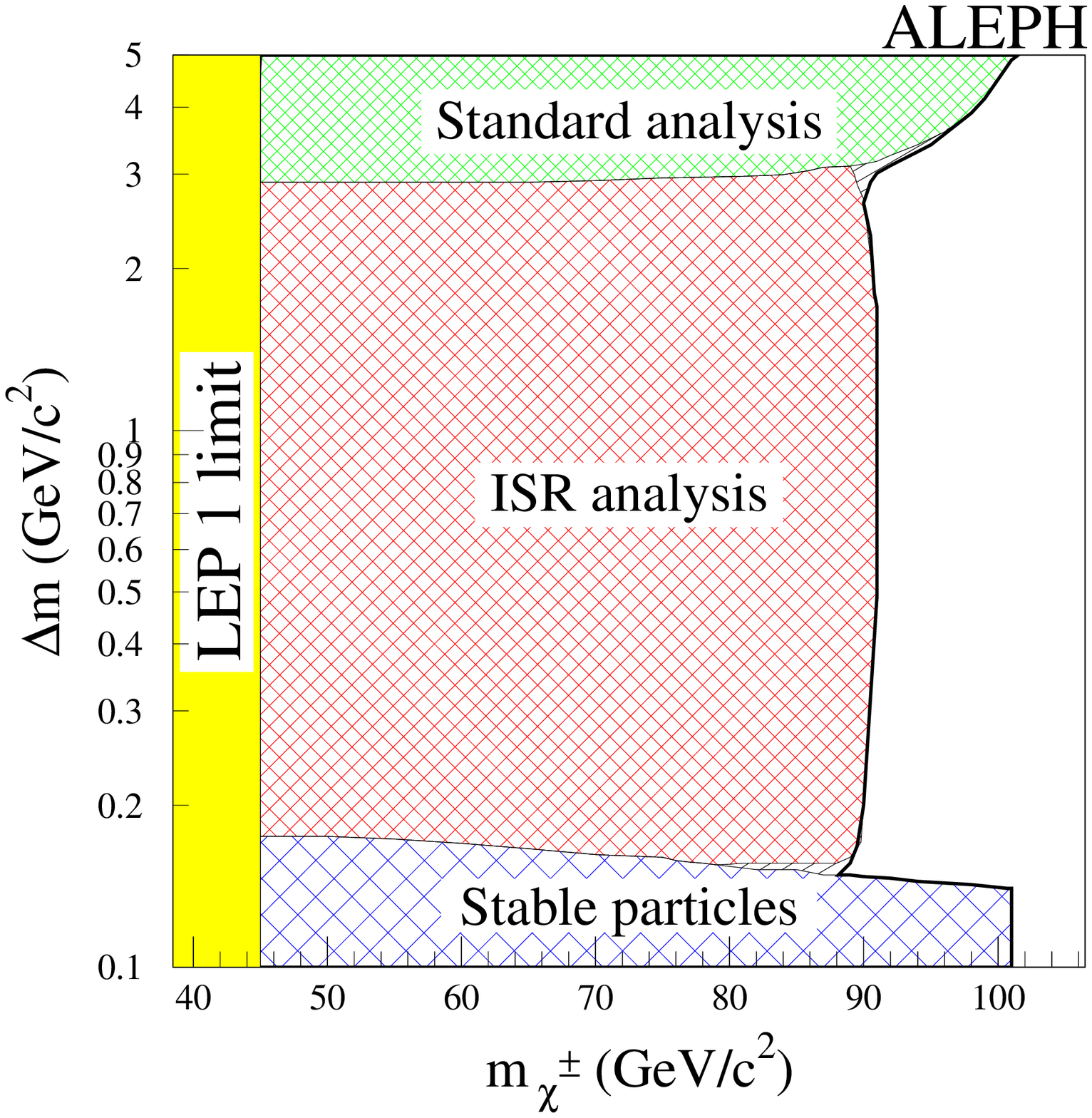}
\def\baselinestretch{1.2}
\caption{\footnotesize{Excluded region in the ($m_{\charg},
\Delta m$) plane at 95\% C.L., in the large scalar mass scenario
($m_0=500 \, \gem$) and in the Higgsino region, independently of
$\tan \beta$. The top area is excluded by the standard missing
energy chargino selection~\cite{emiss,emiss_npy} while the bottom area is
excluded by the stable particle search~\cite{stable,stable_npy}. The
intermediate area is excluded by the ISR analysis. The region
excluded by the combination of the three analyses is bounded by
the bold curve. The region excluded at LEP\,1 is also shown.}}
\label{ISRhiggs}
\end{center}
\end{figure}

In the Higgsino region ($\vert\mu\vert \ll M_2$), the chargino
production cross section is smaller than in the gaugino region,
but this effect is compensated by a generically larger chargino
decay length, together with a reduced influence of $\tan\beta$. As
a result, the ISR analysis on its own excludes a region larger
than in the former case, namely $\Delta m$ between 150\,MeV/$c^2$
and 3\,GeV/$c^2$ (Fig.~\ref{ISRhiggs}), with only a slight
dependence on $m_0$. The combination of the three analyses allows
a lower limit on the chargino mass to be set at 88\,GeV/$c^2$ in
the Higgsino region. It was checked with a scan on the $\mu$
parameter that the 88\,GeV/$c^2$ mass limit, obtained in the
gaugino and Higgsino regions, is generally valid for large
sfermion masses.

For small enough sfermion (most importantly slepton) masses, the chargino
lifetime can be substantially reduced, even for very small $\Delta m$
values. As a result, the search for heavy stable particles loses
efficiency for some parameter configurations as soon as $\Delta m
> m_{\rm e}$, while the ISR analysis is inefficient for such small
$\Delta m$ values. Therefore, the only completely
general chargino mass limit is $m_{\rm Z}/2$, as derived from the
Z width measurement at LEP\,1~\cite{Zwidth}.

In a more constrained version of the MSSM, namely with gaugino-
and sfermion-mass unification, charginos degenerate in mass with
the lightest neutralino are only possible in the deep Higgsino
region. For such large $M_2$ values, all sfermions are heavy. An
absolute chargino mass lower limit of 88\,GeV/$c^2$ therefore
holds within this framework.

\section{Conclusions}

A search for charginos nearly mass degenerate with the lightest
neutralino has been performed using data collected by the ALEPH
detector at LEP at centre-of-mass energies between 189 and
209\,GeV. No excess of candidate events with respect to Standard
Model background predictions was found.

The results have been interpreted in terms of exclusion limits in
the framework of the MSSM; in the heavy sfermion scenario the
absolute 95\%~C.L. chargino mass lower limit is $88~\gem$ for any
$\tan \beta$, irrespective of the chargino field content.

In the light sfermion scenario, no $\Delta m $-independent limit on the 
chargino mass is set from direct searches. Chargino masses up to $
m_{\mathrm{Z}}/2$ are excluded, indirectly, from the measurement
of the Z total width at LEP\,1.

\section*{Acknowledgements}
We wish to congratulate our colleagues from the accelerator
divisions for the successful operation of LEP at high energies. We
would also like to express our gratitude to the engineers and
support people at our home institutes without whom this work would
not have been possible. Those of us from non-member states wish to
thank CERN for its hospitality and support.


\begin{figure}[htbp]
\begin{picture}(160,180)
\put(5,170){\begin{rotate}{-90}{\epsfxsize150mm\epsfbox{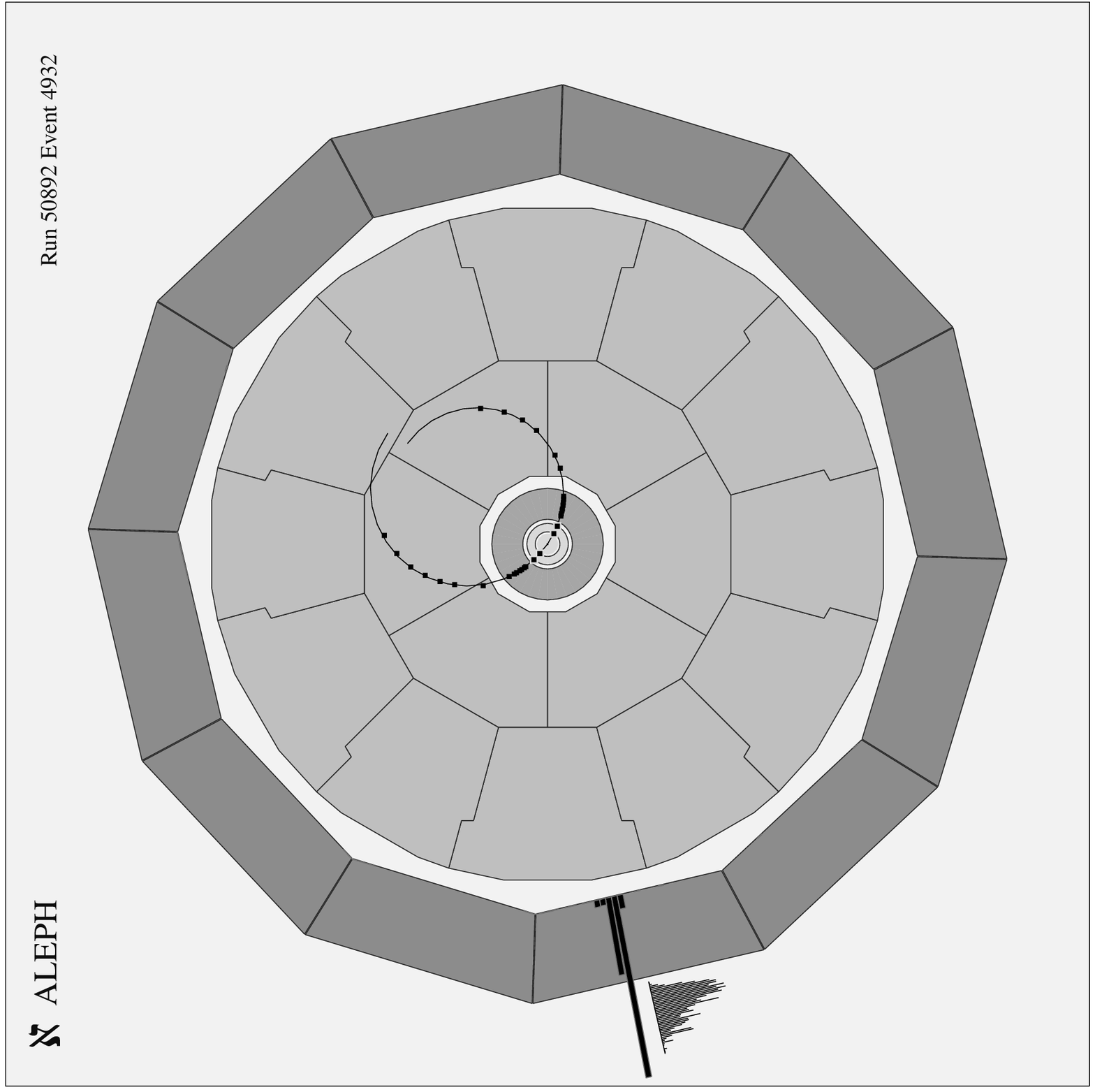}}
\end{rotate}}
\end{picture}
\def\baselinestretch{1.2}
\caption{\footnotesize{Candidate event at $\sqrt{s}=195.5$\,GeV
which contributes to the range $m_{\charg} \leq 84~\gem$. The
reconstructed photon energy is 21\,$\gee$.}}
\label{evento}
\end{figure}

\end{document}